# Through the integration of network pharmacology, molecular docking techniques, and Mendelian randomization analysis, the therapeutic targets and associated biological mechanisms of Gegen Qinlian Decoction in treating Helicobacter pylori were extensively investigated

Ruotong Lu, Xiaozhe Huang, Sihuan Deng, Haikun Du

(Xinjiang Medical University College of pharmacy, Xinjiang Urumq)

**Abstract:** *Objective: This study aimed to utilize network pharmacology, molecular docking techniques, and Mendelian randomization analysis to explore the therapeutic targets and biological mechanisms of Gegen Qinlian Decoction in treating Helicobacter pylori infection and Helicobacter pylori-related gastric cancer, and to identify potential drug targets.*

*Methods: Employing bioinformatics analysis, this research extracted the medicinal components of Gegen Qinlian Decoction, namely Gegen, Huangqin, Huanglian, and Gancao, based on pharmacokinetic criteria from both the TCMSP and HERB databases. Potential disease treatment targets were then retrieved from the DisGeNET and PubChem databases. The STRING database was employed to construct the interaction network of Gegen Qinlian Decoction and Helicobacter pylori infection targets. The "compound-target-disease" network map of Gegen Qinlian Decoction and Helicobacter pylori was constructed using Cytoscape 3.9.1 software. Using both the DAVID enrichment analysis database and the Metascapes gene list analysis website, intersected targets underwent Gene Ontology (GO) and biological pathway (KEGG) enrichment analysis, further exploring the therapeutic effects. Molecular docking and docking result processing were performed using Autodock Tools 1.5.6 and PyMOL 2.5.2 software to investigate the molecular-level relevance between the drug's effective active components and the disease target proteins. Finally, Mendelian randomization analysis was conducted based on the ukb-b-531 sample obtained from UK Biobank (https://www.ukbiobank.ac.uk).*

*Results: From the medicinal components of Gegen Qinlian Decoction, 146 active components and 248 targets were screened. Among these, 66 targets intersected with the target genes of Helicobacter pylori infection and were analyzed as potential therapeutic targets. Based on Cytoscape analysis, molecular docking was conducted between the primary drug active components, namely quercetin, wogonin, kaempferol, and Helicobacter pylori infection-related target genes such as PTGS1, PTGS2, MAPK14. Mendelian randomization analysis revealed several genes significantly associated with Helicobacter pylori infection, including IGF2, PIK3CG, GJA1, and PLAU.*

*Conclusion: The active components in Gegen Qinlian Decoction exhibit high affinity against Helicobacter pylori infection, exerting anti-Helicobacter pylori effects through multiple targets and pathways. Moreover, several potential research targets exist.*

**Key words:** *Helicobacter pylori, Pueraria mirifica pyloria soup, network pharmacology, molecular docking. Mendelian randomization*

.Introduction

**1.1 Research Background**

The bacterium, *Helicobacter pylori*, characterized by its spiral morphology and microaerophilic nature, predominantly colonizes various regions within the stomach and duodenum. Chronic inflammation of the gastric mucosa, albeit mild, is elicited by this microorganism. Such inflammation is implicated in the pathogenesis of both gastric and duodenal ulcers, as well as gastric carcinoma. Studies have revealed that between 67% and 80% of gastric ulcers and as many as 95% of duodenal ulcers can be attributed to the presence of Helicobacter pylori[1][2][3]. The World Health Organization's International Agency for Research on Cancer has classified this bacterium as a Group 1 carcinogen. Epidemiological data suggest that approximately 44% of the global populace is infected, with infection rates in developing nations reaching an alarming 51%[4][5].

### 1.2 Research Questions and Significance

The current conventional quadruple therapy, primarily utilized for the eradication of Helicobacter pylori infection, exhibits suboptimal performance. This regimen is characterized by high antimicrobial resistance, increased adverse reactions, elevated eradication treatment costs, and a propensity for recurrence. As a result, therapeutic strategies leveraging traditional Chinese medicine have emerged as an essential area of investigation[6][7].

### 1.3 Core Concepts

Gegen Qinliantang (GQT) is a distinguished formula originating from the Han Dynasty, specifically derived from the 34th prescription in Zhang Zhongjing's influential work from the late Eastern Han period, the "Treatise on Cold Damage." The formula is meticulously composed of kudzu root (half a pound), roasted licorice (two ounces), baical skullcap (three ounces), and coptis (three ounces)[8]. Traditionally, GQT has been employed in clinical settings for the management of acute enteritis, acute gastritis, and bacterial dysentery. Contemporary animal studies further attest to GQT's efficacy in modulating the gastrointestinal microbiota, thereby rendering protective benefits[9][10]. With this foundational understanding and harnessing the principles of network pharmacology, this research endeavors to dissect the intricate pharmacodynamic interactions of GQT's constituents in combatting *Helicobacter pylori* (HP) infections. Such insights are poised to offer a theoretical scaffold, paving the way for innovative therapeutic strategies targeting *Helicobacter pylori* infections.

### 1.4 Research Methods and Technical Approach

Network pharmacology operates on the premise of the similarities between drugs in terms of structure and function. It also accounts for the myriad interactions between target molecules and bioactive molecules within an organism. By constructing drug-drug and drug-target networks, this discipline predicts the efficacy of drugs and identifies specific efficacies corresponding to particular drugs[11]. In the present study, we initially screened the active constituents of Gegen Qinliantang (GQT). Subsequently, we collated and summarized the targets associated with these active constituents in the treatment of *Helicobacter pylori* infection. Capitalizing on the tools of network pharmacology, Gene Ontology (GO) and Kyoto Encyclopedia of Genes and Genomes (KEGG) pathway enrichment, we explored the targets and pathways pertinent to these active constituents. Drawing from these consolidated findings, molecular docking techniques were employed to analyze the optimal effective components for treating *Helicobacter pylori* infection, aiming to pinpoint the most promising candidates for the development of novel efficacious compounds. The overarching objective of this research is to furnish a theoretical framework for the molecular mechanisms underlying GQT's combat against *Helicobacter pylori* infection.

## 2. Materials and Methods

### 2.1 Collection of Chemical Constituents from Gegen Qinliantang and Screening for Active Compounds

The primary chemical constituents of Gegen Qinliantang, specifically from the four herbal components - kudzu root, licorice, baical skullcap, and coptis - were collated from the TCMSP database (http://tcmspw.com/tcmsp.php) and the HERB database (http://herb.ac.cn/). Compounds were screened based on their pharmacokinetic properties, specifically Absorption, Distribution, Metabolism, and Excretion (ADME) parameters. Criteria for inclusion encompassed compounds with an oral bioavailability (OB) of ≥30% and drug likeness (DL) of ≥0.18. Compounds not meeting these criteria were excluded. The resultant compounds are characterized by favorable drug-likeness and high oral bioavailability and were subsequently cataloged for further analysis[12].

### 2.2 Prediction of Active Compound Targets and Disease Targets

After a meticulous screening of active compounds, they were sequentially introduced into the PubChem database (https://pubchem.ncbi.nlm.nih.gov) for the identification of associated target sites. Subsequent to this, the derived targets were uploaded to the Uniprot database (https://www.uniprot.org) for conversion, with "Homo sapiens" designated as the reference organism. This process enabled the acquisition of targets specifically associated with the active compounds present in Gegen Qinliantang. Additionally, the DisGeNET database (https://www.disgenet.org/) was utilized, employing "Helicobacter pylori (26695)" as the search criterion, to identify pertinent disease targets, which are set to be the focal point of subsequent research endeavors.

### 2.3 Intersection of Active Compound Targets and Disease Targets

The identified targets associated with the active compounds of Gegen Qinliantang and those related to *Helicobacter pylori* infection were imported into the Venny 2.1.0 tool (https://bioinfogp.cnb.csic.es/tools/venny/index.html). A Venn diagram was constructed to establish the intersection of the active compound targets and disease targets. The intersecting targets were subsequently extracted, serving as potential targets for further analysis and operations.

### 2.3 Intersection of Active Compound Targets and Disease Targets

The selected targets from the active compounds of Gegen Qinliantang and those associated with Helicobacter pylori infection were imported into the Venny 2.1.0 tool (https://bioinfogp.cnb.csic.es/tools/venny/index.html). A Venn diagram was subsequently constructed to identify the overlap between the compound-related targets and the disease-specific targets. These overlapping targets were then extracted, serving as potential targets for subsequent investigations.

### 2.4 Construction of the Protein-Protein Interaction (PPI) Network

The intersecting targets identified were imported into the STRING 11.5 database (https://cn.string-db.org/), with "Homo sapiens" selected as the organism of interest for the analysis. A minimum confidence level of 0.700 was set for the Protein-Protein Interaction (PPI) assessment. Based on this confidence threshold, nodes with lower scores or those that stood independently were omitted, leading to the construction of a robust protein interaction network (PPI). This constructed PPI network was then extracted for further examination.

### 2.5 Establishment of the Traditional Chinese Medicine-Active Ingredient-Target-Disease Network

The discerned targets from the active compounds of Gegen Qinliantang, along with those associated with *Helicobacter pylori* infection, were imported into the Cytoscape 3.9.1 software (https://cytoscape.org). Within this platform, a comprehensive network encompassing traditional Chinese medicine, active ingredients, targets, and disease was constructed. Subsequent to this network creation, the relationships between the molecules were analyzed to determine the degree of association, facilitating the identification of core targets and principal compounds.

### 2.6 Biological Analysis

The identified targets were subjected to biological analysis using the DAVID website (https://david.ncifcrf.gov/). A threshold was set with $P<0.05$, and the resultant data were exported to Excel. This data, in conjunction with the Metascape database (https://metascape.org), was employed to submit genes, with "H. sapiens" selected as the species for enrichment calculations. Following the acquisition of results, visual representations were generated in the form of Enrichment Networks, Protein Complex Networks, Enrichment dot bubbles, and Gene Ontology visualizations. These visualized data sets facilitated an in-depth analysis of the outcomes.

### 2.7 Molecular Docking Procedure

The 3D molecular structure of the target was retrieved from the PDB database (https://www.rcsb.org/). Small molecules were searched by their CAS numbers on the PDB database homepage, and complexes containing these ligands were saved. Using the AutoDockTools 1.5.6 software (https://autodock.scripps.edu/), processes such as hydrogen addition, water removal, and ligand removal were performed. Semi-flexible docking procedures were adopted, setting the number of runs to 50. Binding site parameters, binding energies, and binding sites were obtained. The docking results were then exported in the pdbqt format.

Subsequently, Open Babel GUI v2.3.1 (https://openbabel.org) was employed to convert the pdbqt format to pdb format. The processed protein and small molecule pdb files were imported into PyMOL 2.5.2 (www.pymol.org) for visualization. Based on the binding energies at different binding sites, sites exhibiting superior binding activity were identified and analyzed.

**2.8 Mendelian Randomization-Based Study:**

To investigate the relationship between the relevant genes and HP, we initially utilized the target genes identified and subsequently employed the Mendelian Randomization (MR) method to assess the association between these target genes and the disease. MR is a statistical approach that facilitates the evaluation of causal relationships between exposure and outcomes. Our exposure data were sourced from the UK Biobank (https://www.ukbiobank.ac.uk), specifically from the ukb-b-531 sample dataset, which represents a comprehensive, large-scale national biorepository.

In our Mendelian randomization analysis, we employed the TwoSampleMR （https://github.com/MRCIEU/TwoSampleMR） package. We set the threshold for the minimum LD $R^2$ value at 0.8 to ensure the robustness of the instrumental variables. For aligning palindromes, a minor allele frequency (MAF) threshold of 0.3 was established. We conducted multiple MR analyses using various methods, including Wald ratio, MR Egger, Weighted median, Inverse variance weighted, and Weighted mode. To mitigate potential biases and enhance the validity of our results, we rigorously excluded SNPs in linkage disequilibrium

## 3. Results

### 3.1 Selection of Efficacious Components in Gegen Qinliantang

Utilizing the TCMSP and HERB databases, searches were conducted using the pharmacologically active ingredients of Gegen Qinliantang, specifically "Pueraria (Ge Gen)", "Licorice (Gan Cao)", "Scutellaria (Huang Qin)", and "Coptis (Huang Lian)", as keywords[12]. Compounds were screened based on pharmacokinetic criteria, with an oral bioavailability (OB) of ≥30% and a drug likeness (DL) value of ≥0.18. In total, 146 active compounds were identified across these four ingredients: 4 from Pueraria, 92 from Licorice, 36 from Scutellaria, and 14 from Coptis. A detailed representation of some of these active components can be found in Figure 1.

| No. | Description | MyList | Gene ID | No. | Description | MyList | Gene ID |
|---|---|---|---|---|---|---|---|
| 1 | prostaglandin-endoperoxide synthase 2 | PTGS2 | 5743 | 34 | erb-b2 receptor tyrosine kinase 2 | ERBB2 | 2064 |
| 2 | C-X-C motif chemokine ligand 8 | CXCL8 | 3576 | 35 | mitogen-activated protein kinase 14 | MAPK14 | 1432 |
| 3 | interleukin 6 | IL6 | 3569 | 36 | AKT serine/threonine kinase 1 | AKT1 | 207 |
| 4 | interleukin 1 beta | IL1B | 3553 | 37 | NAD(P)H quinone dehydrogenase 1 | NQO1 | 1728 |
| 5 | interferon gamma | IFNG | 3458 | 38 | ATP binding cassette subfamily G member 2 (Junior blood group) | ABCG2 | 9429 |
| 6 | interleukin 10 | IL10 | 3586 | 39 | caveolin 1 | CAV1 | 857 |
| 7 | BCL2 apoptosis regulator | BCL2 | 596 | 40 | neutrophil cytosolic factor 1 | NCF1 | 653361 |
| 8 | tumor protein p53 | TP53 | 7157 | 41 | secretory leukocyte peptidase inhibitor | SLPI | 6590 |
| 9 | tumor necrosis factor | TNF | 7124 | 42 | superoxide dismutase 1 | SOD1 | 6647 |
| 10 | signal transducer and activator of transcription 3 | STAT3 | 6774 | 43 | cyclin D1 | CCND1 | 595 |
| 11 | mitogen-activated protein kinase 1 | MAPK1 | 5594 | 44 | caspase 3 | CASP3 | 836 |
| 12 | interleukin 2 | IL2 | 3558 | 45 | RELA proto-oncogene, NF-kB subunit | RELA | 5970 |
| 13 | secreted phosphoprotein 1 | SPP1 | 6696 | 46 | RB transcriptional corepressor 1 | RB1 | 5925 |
| 14 | myeloperoxidase | MPO | 4353 | 47 | caspase 8 | CASP8 | 841 |
| 15 | epidermal growth factor receptor | EGFR | 1956 | 48 | FOS like 1, AP-1 transcription factor subunit | FOSL1 | 8061 |
| 16 | C-reactive protein | CRP | 1401 | 49 | fatty acid synthase | FASN | 2194 |
| 17 | epidermal growth factor | EGF | 1950 | 50 | E2F transcription factor 1 | E2F1 | 1869 |
| 18 | vascular endothelial growth factor A | VEGFA | 7422 | 51 | glycogen synthase kinase 3 beta | GSK3B | 2932 |
| 19 | transforming growth factor beta 1 | TGFB1 | 7040 | 52 | gap junction protein alpha 1 | GJA1 | 2697 |
| 20 | Fos proto-oncogene, AP-1 transcription factor subunit | FOS | 2353 | 53 | activator of HSP90 ATPase activity 1 | AHSA1 | 10598 |
| 21 | serpin family E member 1 | SERPINE1 | 5054 | 54 | cytochrome P450 family 3 subfamily A member 4 | CYP3A4 | 1576 |
| 22 | nuclear receptor subfamily 1 group I member 2 | NR1I2 | 8856 | 55 | matrix metallopeptidase 9 | MMP9 | 4318 |
| 23 | C-C motif chemokine ligand 2 | CCL2 | 6347 | 56 | matrix metallopeptidase 1 | MMP1 | 4312 |
| 24 | glutathione S-transferase mu 1 | GSTM1 | 2944 | 57 | phosphatidylinositol-4,5-bisphosphate 3-kinase catalytic subunit gamma | PIK3CG | 5294 |
| 25 | peroxisome proliferator activated receptor gamma | PPARG | 5468 | 58 | nitric oxide synthase 2 | NOS2 | 4843 |
| 26 | MYC proto-oncogene, bHLH transcription factor | MYC | 4609 | 59 | hypoxia inducible factor 1 subunit alpha | HIF1A | 3091 |
| 27 | prostaglandin-endoperoxide synthase 1 | PTGS1 | 5742 | 60 | heat shock protein 90 alpha family class A member 1 | HSP90AA1 | 3320 |
| 28 | MCL1 apoptosis regulator, BCL2 family member | MCL1 | 4170 | 61 | intercellular adhesion molecule 1 | ICAM1 | 3383 |
| 29 | BCL2 associated X, apoptosis regulator | BAX | 581 | 62 | insulin like growth factor 2 | IGF2 | 3481 |
| 30 | peroxisome proliferator activated receptor delta | PPARD | 5467 | 63 | heat shock protein family B (small) member 1 | HSPB1 | 3315 |
| 31 | ornithine decarboxylase 1 | ODC1 | 4953 | 64 | heme oxygenase 1 | HMOX1 | 3162 |
| 32 | plasminogen activator, urokinase | PLAU | 5328 | 65 | Fas ligand | FASLG | 356 |
| 33 | sirtuin 1 | SIRT1 | 23411 | 66 | interleukin 4 receptor | IL4R | 3566 |

**Figure 1**

## 3.2 Prediction of Targets for Active Compounds and Disease Targets Related to Helicobacter pylori Infection

Active compound targets from the 146 constituents of Gegen Qinliantang were retrieved from the TCMSP database. After screening, a total of 2,185 potential target genes were identified. After eliminating duplicates and subsequent validation, 248 unique target genes were obtained. These 248 targets were then converted within the Uniprot database, specifically focusing on the Homo sapiens (human) organism for conversion, thereby obtaining their respective gene identifiers. This process yielded 236 potential gene targets post-screening. Utilizing the DisGeNET database and employing "Helicobacter pylori" as the search term, related gene targets for Helicobacter pylori infection were identified. After removing duplicates, a total of 593 target genes were delineated.

## 3.3 Intersection of Active Compound Targets and Disease Targets Associated with Helicobacter pylori Infection

The 248 screened active compound targets and the 593 disease targets associated with Helicobacter pylori infection were imported into the online tool Venny 2.1.0 (https://bioinfogp.cnb.csic.es/tools/venny/index.html) to generate a Venn diagram, illustrating the intersection between the active compound targets and disease targets. From this process, 66 overlapping targets were identified. These 66 shared targets were subsequently highlighted as the key targets for further study. Notably, the intersecting targets between Gegen Qinliantang and Helicobacter pylori infection constituted 8.7% of the total targets. Refer to Figure 2 for a detailed representation.

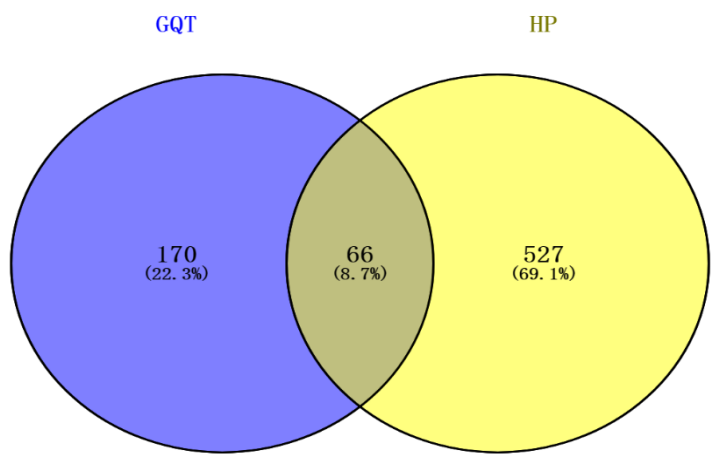

**Figure 2**

## 3.4 Protein-Protein Interaction Network of Key Targets

The previously identified 66 key target proteins were imported into the STRING 11.5 database (https://cn.string-db.org/) for protein-protein interaction (PPI) analysis. The organism of interest was set to "Homo sapiens." A minimum confidence level of 0.700 was established, and based on this confidence threshold, nodes with lower significance or that stood independently were excluded. This facilitated the construction of a protein-protein interaction network (PPI), as depicted in Figure 3. Upon analysis of the network, a total of 66 nodes and 454 edges were identified, with an average node degree of 13.8 and an average local clustering coefficient of 0.546.

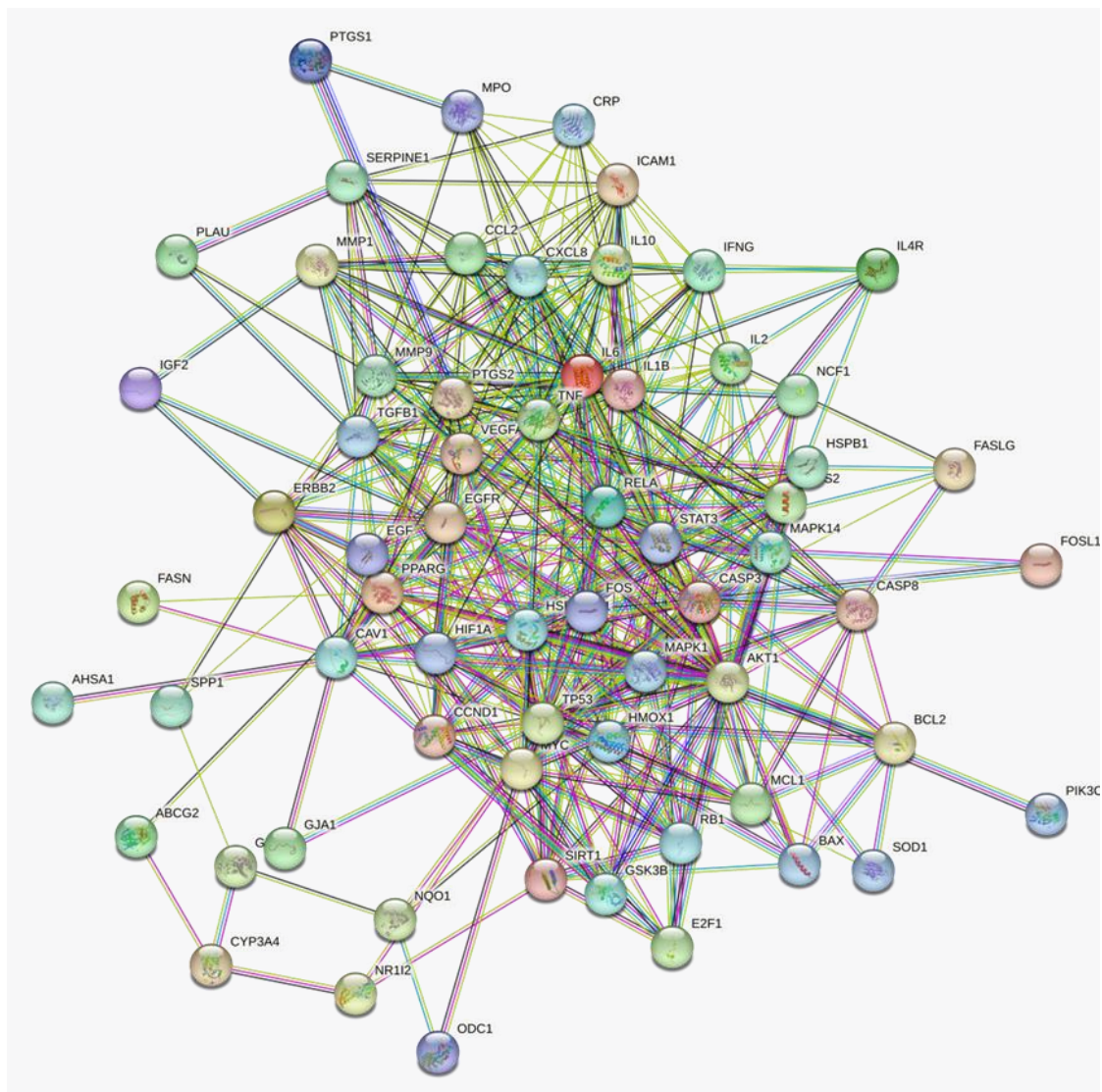

**Figure 3**

### 3.5 Construction of the Traditional Chinese Medicine (TCM)-Active Component-Target-Disease Network

The 66 target proteins, identified from the previous steps, were subjected to the UniProt database (https://www.uniprot.org/) to retrieve their corresponding gene names. Alongside, the 146 principal active components of the medicinal compounds extracted from the TCMSP database were imported into the Cytoscape 3.9.1 software. This facilitated the creation of the TCM-Active Component-Target-Disease network. Upon analysis (as illustrated in Figure 4), a total of 118 nodes and 419 edges were identified. In the network diagram, triangles represent Helicobacter pylori infections, rectangles signify targets, ellipses denote active drug compounds, diamonds represent the drug, and octagons symbolize Gegen Qinliantang.

Based on the number of connections linked to the disease target nodes, four core targets emerged with the highest degree of association: PTGS2 (93 connections), NOS2 (73 connections), PTGS1 (59 connections), and MAPK14 (39 connections). Similarly, based on the connection count of the active drug compound nodes, three primary active compounds stood out for their high degree of association: quercetin (24 connections), wogonin (15 connections), and kaempferol (8 connections), with the numbers in parentheses indicating the degree of relatedness. Consequently, the identified four core disease targets and three principal active compounds were chosen as the prime candidates for molecular docking studies.

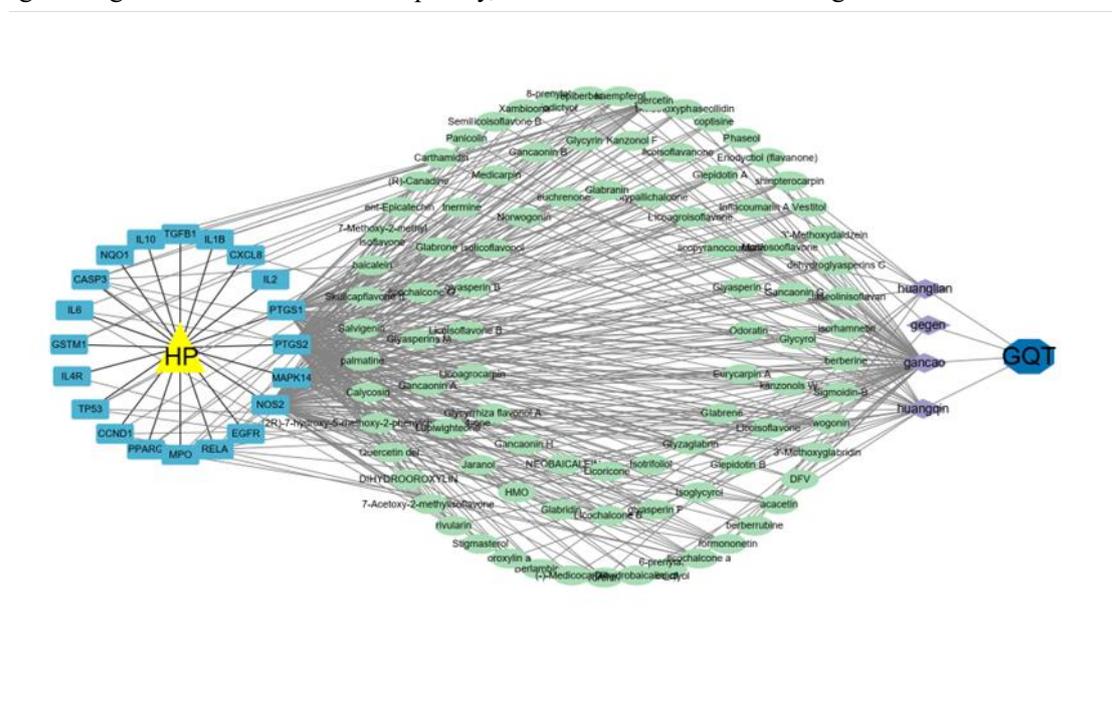

**Figure 4**

### 3.6 Biological Analysis

Biological analysis of the identified targets was carried out using the DAVID enrichment analysis database (https://david.ncifcrf.gov/). A threshold was set at P<0.05, and the resulting data were exported to Excel. This data was subsequently uploaded to the Metascape gene annotation tool (https://metascape.org). Genes were submitted with "H.sapiens" selected as the target organism for enrichment calculations. The options for GO Molecular Functions, GO Biological Processes, and GO Cellular Components were selected for further analysis. Upon completion, results were sorted based on their Z-score, with the top 10 pathways selected for visualization, resulting in a bar chart for GO analysis, as depicted in Figure 5. In this context, BP, CC, and MF respectively denote biological processes, cellular components, and molecular function categories. For pathway analysis, the KEGG Pathway option was checked in the Metascape database (https://metascape.org). The resulting data, after analysis, were sorted by P-value, with the top twenty pathways based on the highest P-values chosen for examination. Counts for each pathway were extracted, and a visualization was created in the form of a KEGG pathway bubble chart, as illustrated in Figure 6.

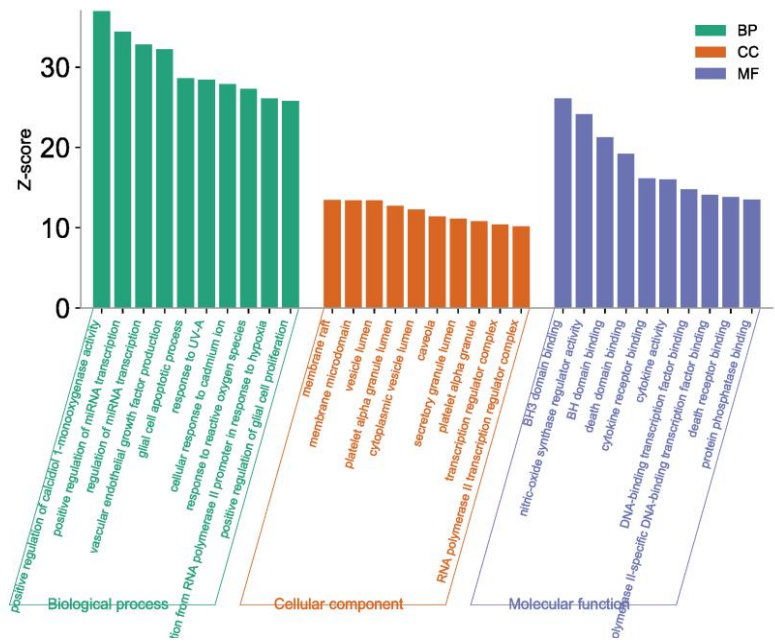

**Figure 5**

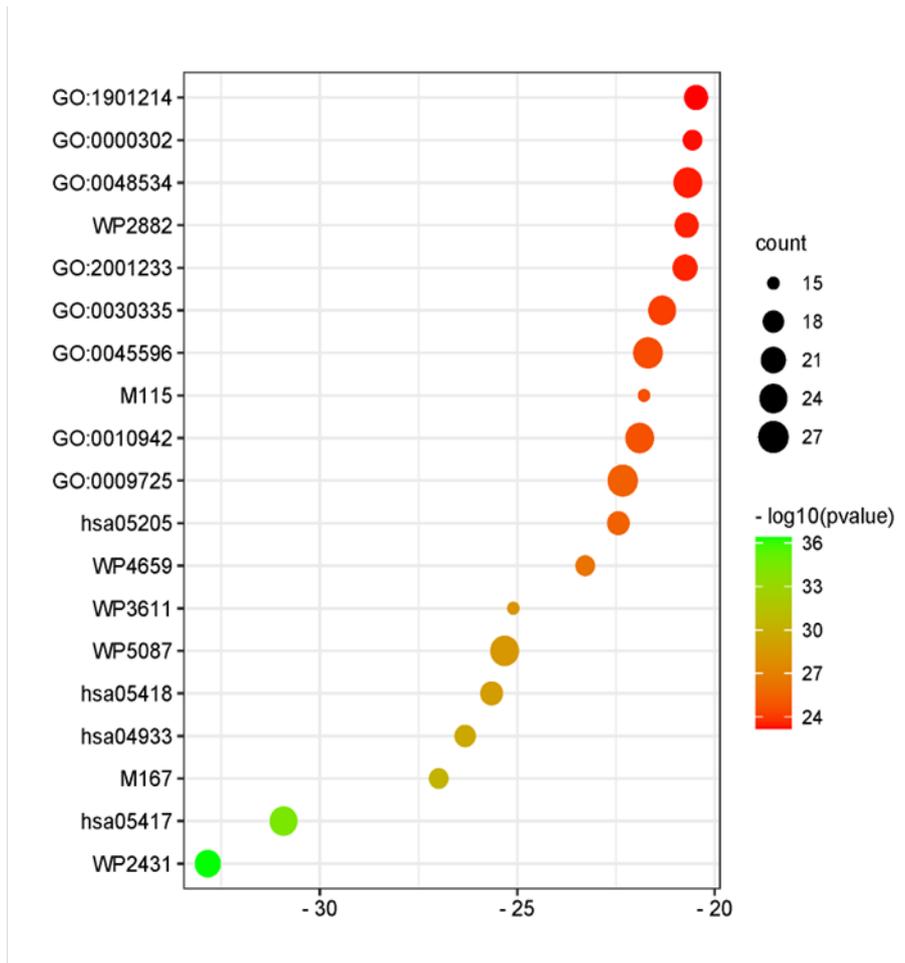

**Figure 6**

### 3.7 Molecular Docking

From the Protein Data Bank (PDB, https://www.rcsb.org/), molecular 3D structures of four core disease targets and three core active compounds were retrieved and downloaded, exported in pdb format. Small molecule CAS numbers were searched within the PDB to identify complexes containing these ligands, which were then saved in mol2 format. Using AutoDockTools 1.5.6 software (https://autodock.scripps.edu/), hydrogenation, water removal, and ligand removal processes were performed, followed by a semi-flexible docking approach set for 50 iterations. Once the rendering was complete, the docking results were acquired, with the scenarios displaying the highest binding affinity saved and then exported in pdbqt format.

These pdbqt formatted results were converted into pdb format using the Open Babel GUI v2.3.1 software (https://openbabel.org) and subsequently visualized in the PyMOL 2.5.2 software (www.pymol.org), as shown in Figure 8. The visualization was adjusted based on binding site parameters, binding energy comparisons, and binding sites. The binding energy of different sites was used to assess the quality of the interaction. Specifically, a binding energy less than -7.0 kcal/mol indicates the presence of binding activity. If the binding energy is less than -3.5 kcal/mol[13], it suggests a favorable binding affinity between the ligand and the receptor1, allowing for the identification of binding sites with better activity, as depicted in Figure 7.

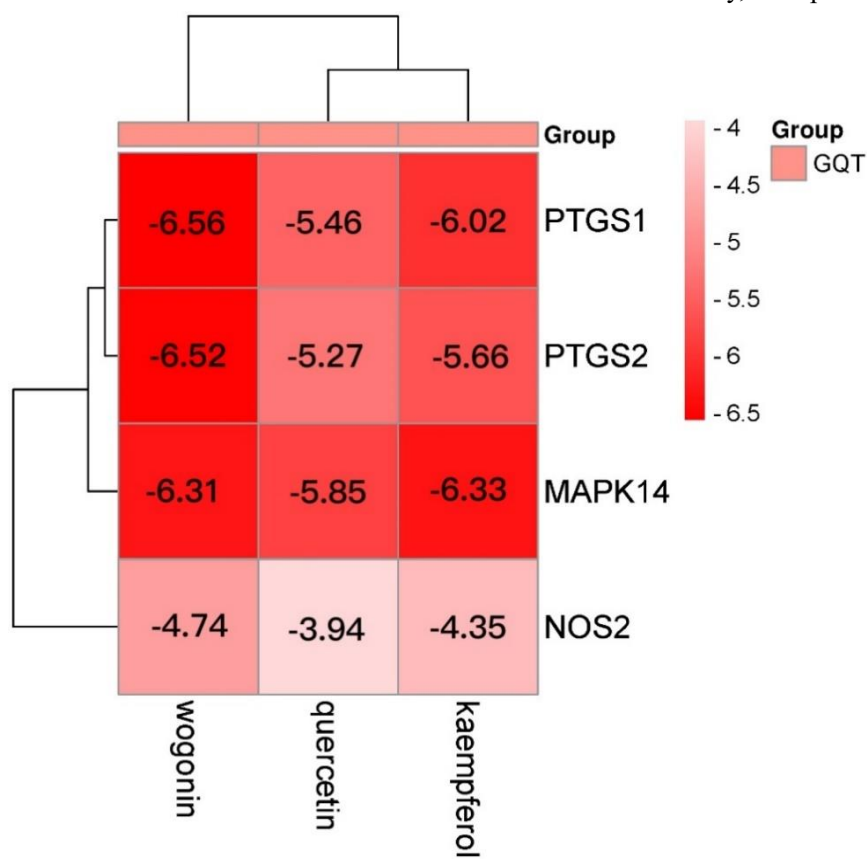

**Figure 7**

10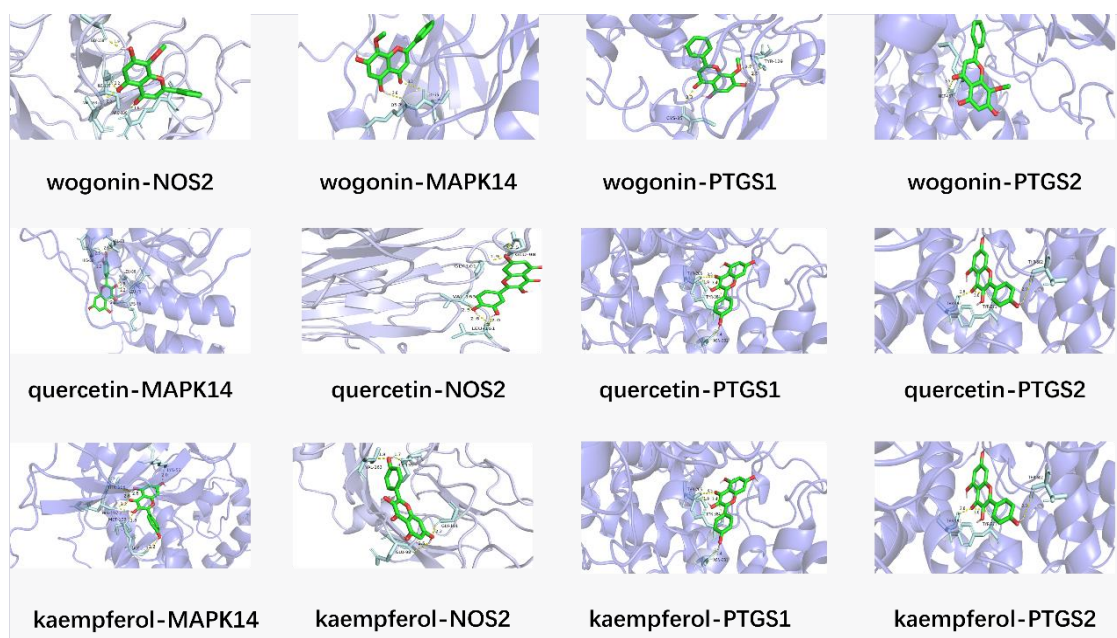

Figure 8

## 3.8 Results and Discussion

Based on the results of Mendelian randomization analysis, this study employed three distinct methods to meticulously examine the relationship between specific genes and Helicobacter pylori infection (HP)[14][15]. In Figure 9, the effect size (b value) for each gene and its corresponding 95% confidence interval are illustrated. Genes are ordered by their effect sizes. Each gene is represented by a horizontal line indicating its 95% confidence interval, with a dot denoting its effect size. Genes with statistically significant associations are highlighted in red, as depicted in Figure 10.

Upon observation, it was found that the 95% confidence intervals of most genes encompassed zero. However, certain genes, such as IGF2, PIK3CG, GJA1, and PLAU, had confidence intervals that did not cover zero, suggesting a potential causal association with the sample. Specifically:

IGF2: Exhibits a negative effect size (b value of -0.000990) with a 95% confidence interval of [-0.001926, -0.000053] and a p-value of 0.03832. This implies a statistically significant negative association between the IGF2 gene and the study outcomes.

PIK3CG: Has an effect size of 0.001106, a 95% confidence interval of [0.000113, 0.002099], and a p-value of 0.02903, indicating a positive statistically significant relationship between the PIK3CG gene and the outcomes.

GJA1: Displays an effect size of 0.001308, a 95% confidence interval of [0.000117, 0.002499], and a p-value of 0.03139, further showcasing a positive statistically significant association.

PLAU: Possesses the largest effect size at 0.002022, with a 95% confidence interval of [0.000458, 0.003586] and a p-value of 0.01128, denoting a strong positive association between the PLAU gene and the study results.

Genes highlighted in red, whose 95% confidence intervals did not surpass zero, showed statistically significant associations with the study outcomes. These genes might biologically relate to the study outcomes. However, further experiments and studies are required to ascertain the biological relevance of these associations.

For enhanced visualization and analysis, a funnel plot was constructed based on the data, as seen in Figure 11. Funnel plots are typically used to display heterogeneity among studies, with the effect size (b values) on the x-axis and their standard errors (se) on the y-axis. The plot should ideally resemble an inverted funnel, where larger studies (with smaller standard errors) are positioned at the top, and smaller studies (with larger standard errors) are dispersed at the bottom. Any significant deviation from this symmetrical shape might suggest publication bias or other forms of bias. In the figure, there wasn't an apparent deviation from this symmetry, indicating potential absence of publication or other biases. The red dots (highlighted genes) were located at the mid and top sections of the funnel, suggesting they have medium to small standard errors. These points have significant effect sizes compared to others and relatively small standard errors, further reinforcing the credibility of their association with the study outcomes. Most data points clustered near an effect size of zero, aligning with our observation in the forest plot that most genes might not have a statistically significant association with the study outcomes.

Figure 12 showcases the relationship between effect size (b values) and -log10(p values). Using -log10(p values) on the y-axis allows for a clearer visualization of smaller p-values. In the figure, higher points represent smaller p-values, indicating statistically significant associations. A horizontal dashed line represents the threshold of p-value at 0.05, a common significance level. Points below this line generally are not considered statistically significant. In the figure, many points lie above this line, indicating their p-values are less than 0.05. The red points (highlighted genes) all have smaller p-values, further emphasizing their statistically significant association with the study outcomes.

12 :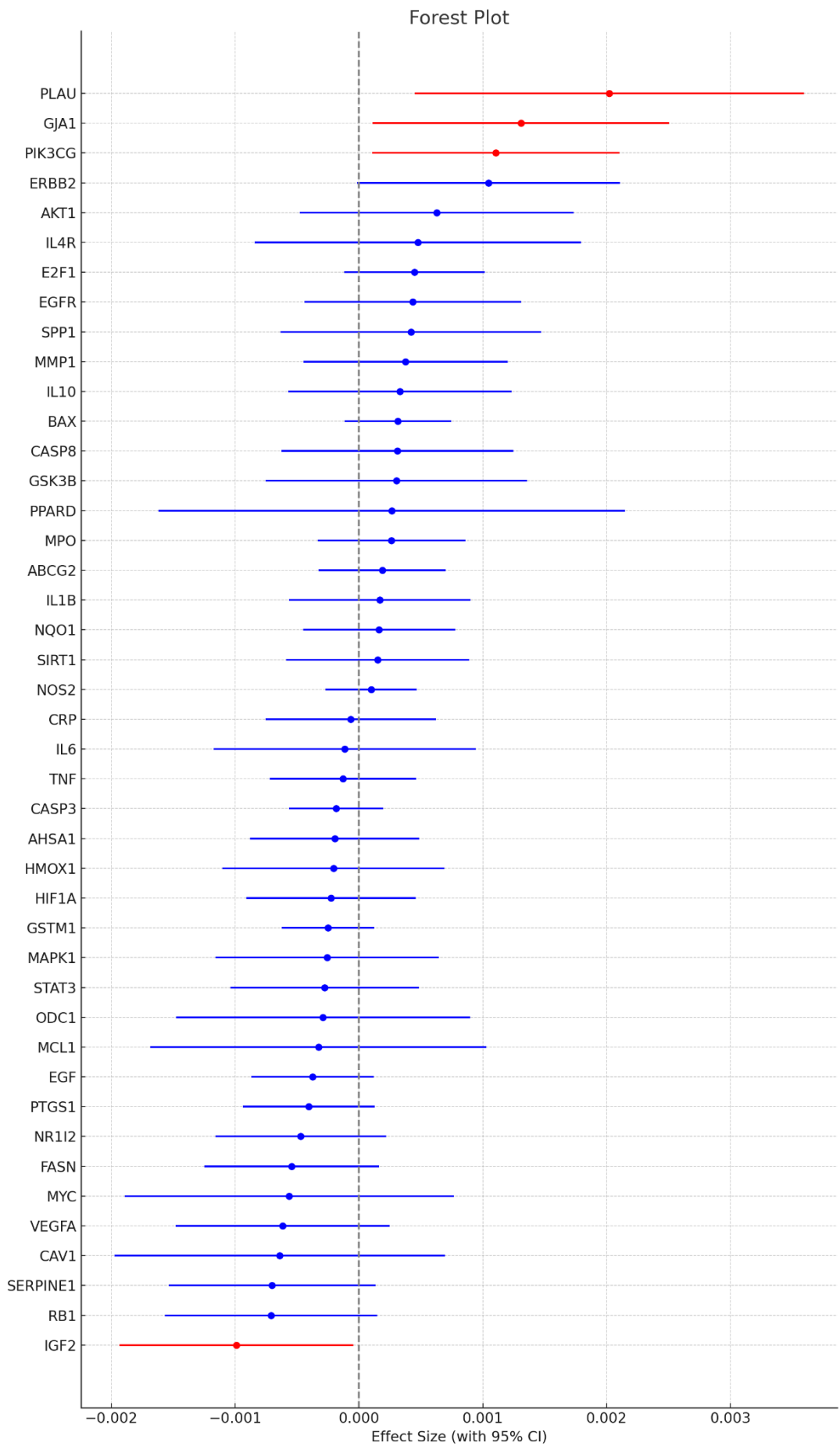

**Figure 9**

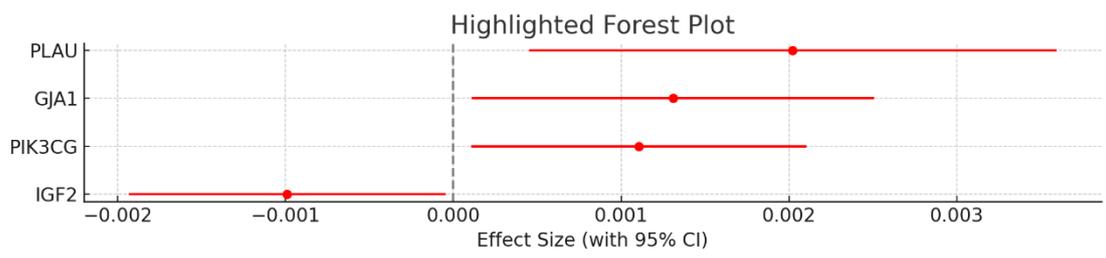

**Figure 10**

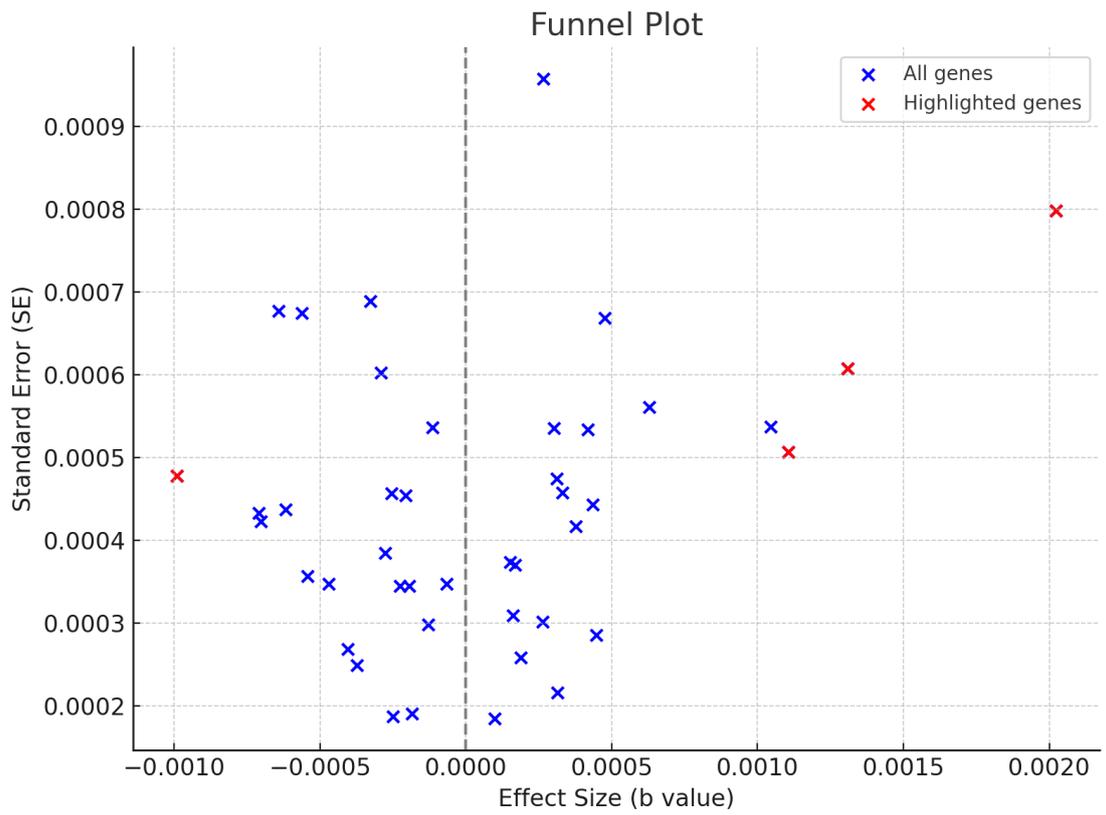

**Figure 11**

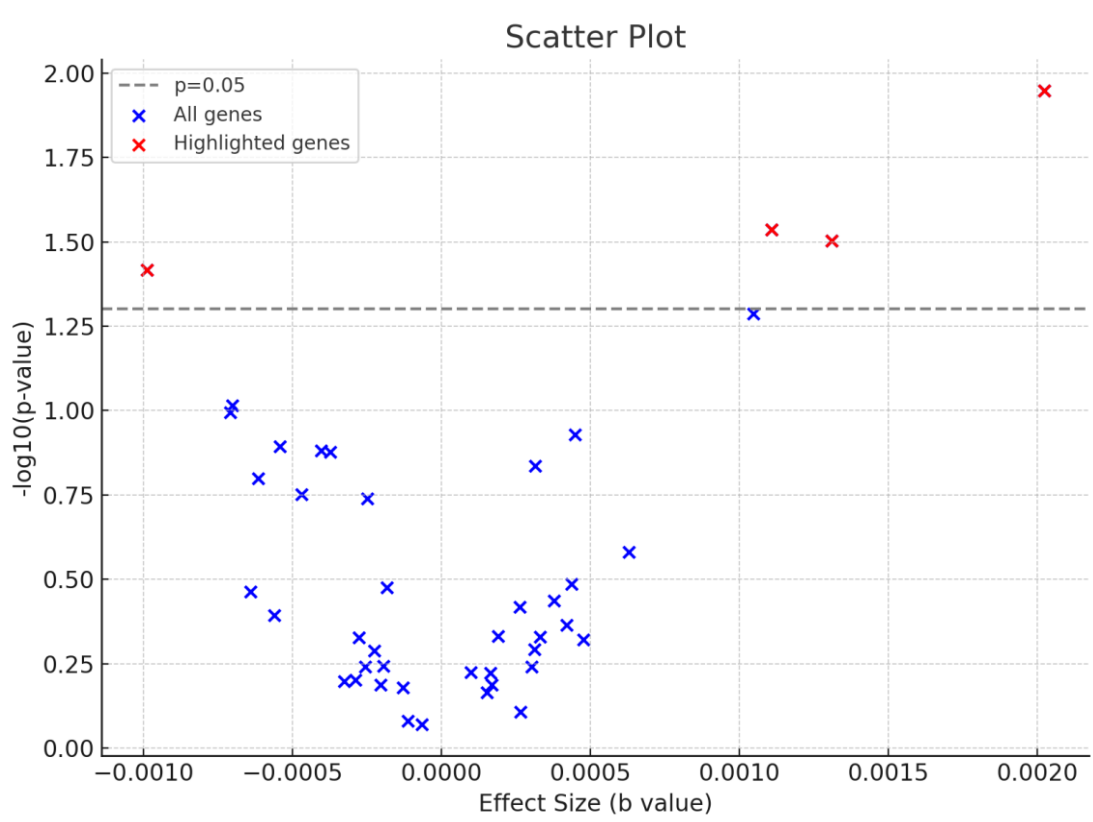

**Figure 12**

## 4. Results and Discussion

In this study, network pharmacology and molecular docking methods were employed to analyze the potential mechanism of Gegen Qinlian Decoction in the treatment of Helicobacter pylori infection. Distinct from conventional therapeutic methods, Gegen Qinlian Decoction is characterized by its safety and minimal toxic side effects. A total of 66 active components were identified in this medicine, which may exert their effects through multiple targets, including PTGS2, PTGS1, MAPK14, and NOS2, among others. The decoction potentially acts through diverse mechanisms to combat Helicobacter pylori infection.

Based on the results of the GO biological analysis, the active components of Gegen Qinlian Decoction act on several signaling pathways. In biological processes, they regulate various events such as the regulation of miRNA transcription, positive regulation of miRNA transcription, and vascular endothelial growth factor production. These actions promote the repair of gastric epithelial cell damage caused by Helicobacter pylori infection[16][17]. At the cellular component level, they influence structures such as the membrane raft, membrane microdomain, vesicle lumen, and platelet alpha granule lumen, intervening in the damage caused by the pathogenic factors produced by Helicobacter pylori[19]. In terms of molecular function, they modulate BH3 domain binding, nitric oxide synthase regulatory activity, and cytokine receptor binding, thereby hindering cell death caused by infection. This modulation affects the inhibitory effects of Helicobacter pylori's Vac A toxin on cell repair, impedes its induced apoptosis, interferes with gastric mucosal nitric oxide synthase activity and cytokine receptor binding, and impacts gastritis induced by Helicobacter pylori infection[20][21].

KEGG enrichment analysis revealed that Gegen Qinlian Decoction participates in regulatory pathways such as the modulation of neuronal death (GO:1901214), response to reactive oxygen species (GO:0000302), and hematopoietic or lymphoid organ development (GO:0048534). By modulating these pathways, the decoction further aids in the repair of gastric epithelial cell damage caused by infection.

Molecular docking results indicate that the active components within Gegen Qinlian Decoction possess robust binding affinities. Specifically, quercetin, wogonin, and kaempferol demonstrate strong binding properties, potentially counteracting Helicobacter pylori infection. Among the potential targets, PTGS2, PTGS1, MAPK14, and NOS2 are of notable significance. This research suggests that Gegen Qinlian Decoction can combat Helicobacter pylori infection through multiple components, targets, and pathways.

This study utilized network pharmacology and molecular docking methods to delve into the molecular mechanisms of Gegen Qinlian Decoction in treating Helicobacter pylori infection. Based on the constructed "Traditional Chinese Medicine-Active Ingredients-Targets-Disease" network, it was revealed that the 146 active components of the four herbs, Gegen, Gancao, Huangqin, and Huanglian, exert therapeutic effects through the modulation of 30 critical pathways involving 66 targets such as PTGS2, PTGS1, MAPK14, and NOS2. The molecular docking results shed light on the molecular mechanisms through which the herbal medicine treats Helicobacter pylori infection, encompassing biological functions, cellular components, molecular functions, enrichment analysis, and molecular docking. Most genes associated with HP infection did not appear to have a significant statistical correlation with the ukb-b-531 samples. This might suggest that the role of these genes in the biological processes related to Helicobacter pylori infection does not directly relate to the physiological or disease states represented by the ukb-b-531 samples.

However, several genes, such as IGF2, PIK3CG, GJA1, and PLAU, were identified to have a statistically significant association with the ukb-b-531 samples. These genes might serve as critical links between the biological mechanisms of Helicobacter pylori infection and the physiological or disease states represented by the ukb-b-531 samples.

It's worth noting that, although our analysis provides compelling evidence of the association between these genes and the ukb-b-531 samples, further experimental validation is required to ascertain whether these observed associations have biological significance.